\documentclass[a4paper]{article}
\usepackage{epsfig, amssymb}
\begin{document}
\title{\huge{\textbf{Techniques for solving bound state problems}}}
\author{M. van Iersel, C.F.M. van der Burgh, and B.L.G. Bakker}
\date{\today}
\maketitle

\section{Introduction}

The Hamiltonian of a system of two scalar particles with mass $m$ in
a local potential is in the relativistic situation given by
\begin{equation}
H = 2 \sqrt{m^2 + p^2} + V(r)
\label{hamrel}
\end{equation}
We are interested in solving the Schr\"odinger equation for binding
and confining potentials, that are associated with particle exchange.

There are several methods for obtaining bound states. Probably the
simplest way is to directly solve the differential equation in
configuration space. Unfortunately we cannot use this method in the
relativistic case, since we have to deal with a term  that contains
a square root of the differential operator.

Another possible method is to directly discretize the integral equation
in momentum space. This comes down to solving an equation of the form
\begin{equation}
E \psi_{i} = \sum_{j} K_{ij} \psi_{j} \, .
\end{equation}
A third method makes use of the variational principle for the energy
and expands the eigenstates in terms of basis functions ($\psi = \sum_{i}
a_{i} \phi_{i}$). It turns out to be advantageous to be able to
calculate the kinetic part of the Hamiltonian in momentum space
and the potential part in configuration space.

In this report we will calculate the bound states with both methods.
The issues we concentrate on are ease of numerical implementation,
accuracy and stability. Both methods are less easy to implement and
calculations take up more time compared to the method of directly
solving the differential equation. To compare the accuracy of the
two methods, we first need to know whether our codes are correct.
To check our codes we have made use of several potentials for which
the bound state energies and wave functions are known explicitly in
the nonrelativistic case. In section 2 we will present these
potentials.  In the next section we will look in more detail at
the numerical implementation of the methods. In section 4 we will
give some numerical results in the nonrelativistic case and in
the next section we will look at some physical interesting cases.
Finally we will give some conclusions in the last section.

\section{Exact solutions of potentials in the nonrelativistic case}

There are several potentials for which the bound state energies and wave
functions are known in the nonrelativistic situation. We have used
three of these potentials to check our codes. They are the Coulomb
potential, the linear potential and the Hulth\'en potential. Note
that two of these potentials have some features in common with
the Yukawa potential: the Coulomb and Hulth\'{e}n potentials
behave like $1/r$ at $r \to 0$; the Hulth\'{e}n potential falls
of like an exponential as $r \to \infty$, almost the same as the
Yukawa potential
\begin{eqnarray}
 V(r) = \frac{\rho}{r} e^{-\mu_{\mathrm{Y}}r} \, .
\label{pot}
\end{eqnarray}
By looking at this potential, Eq.~(\ref{pot}), one realizes
immediately that putting $\mu_{\mathrm{Y}} = 0$ reduces the Yukawa
potential to the Coulomb potential. The energies of the bound
states in the nonrelativistic Coulomb case are known exactly and
given by
\begin{eqnarray}
E_{n} = \frac{m \rho^2}{4n^2}
\end{eqnarray}

\subsection{The linear potential}

In the case of the linear potential ($V(r) = \sigma r$) the eigenvalues
are exactly known for $l = 0$ in the nonrelativistic case and can be
expressed in terms of the zeros of the Airy function \cite{abr&ste}.
The eigenvalues are given by
\begin{eqnarray}
 E_{n} = \left(\frac{\sigma^2}{m}\right)^{1/3} \lambda_{n}
\end{eqnarray}
where $\sigma$ is the strength of the potential and $\lambda_{n}$ is
the n-th zero of the Airy function (see table~\ref{airyzero}).

\subsection{The Hulth\'en potential}

The Hulth\'en potential has the following form
\begin{equation}
 V(r) = \rho \frac{e^{-\mu_{\mathrm{H}}r}}{1 - e^{-\mu_{\mathrm{H}}r}},
\end{equation}
where $\rho$ is the strength of the potential.
The wave functions are known for this potential in the
nonrelativistic case ($l = 0$) and are given by
\begin{equation}
 r \psi (r) = u_{k}(r) = e^{-\alpha_{k}r} (1 - e^{-\mu_{\mathrm{H}}r})
 P_{k}^{(2\alpha_{k}/\mu_{\mathrm{H}},1)} (1 - 2 e^{-\mu_{\mathrm{H}}r}),
\end{equation}
where $\alpha_{k}$ is defined as
\begin{equation}
 \alpha_{k} = \frac{-m\rho -
 (k + 1)^{2}\mu_{\mathrm{H}}^{2}}{2(k + 1)\mu_{\mathrm{H}}} \, ,
\end{equation}
$P_{k}^{(2\alpha_{k}/\mu_{\mathrm{H}},1)}$ is a Jacobi polynomial
\ref{abr&ste}
and the bound state energies are given by
\begin{equation}
 E_{k} = - \frac{\alpha_{k}^{2}}{m}.
\end{equation}

\section{Implementation}

There are different methods to calculate the bound state energies and
wave functions of a system. The methods we present here are chosen for
different reasons; ease of implementation, accuracy, speed, stability
etc. First we will present the different approaches to calculate bound
states. Then we will look in more detail at the methods we use and pay
attention to the stability and accuracy of these methods.

\subsection{Different approaches to calculate bound states}

The first method we mention is the method in which the differential
equation is solved step by step. It is very easy to implement.
The calculations do not take much time, but it is not very
accurate. And unfortunately this method gives problems in the
relativistic case.

Discretizing the integral equation is a more accurate method.
It is less easy to implement and calculations done using this method
take up much more time. This method is not very well suited to all
potentials, since all calculations are done in the momentum space representation
and we may have to deal with singularities in the potential matrix elements. 

The third method we present expands the eigenfunctions in terms of
basis functions and does not have this problem. The two parts of the
Hamiltonian, the kinetic and potential energies respectively, are
calculated in either momentum space or configuration space, whatever is
more convenient.  This method also gives a better accuracy than the
first method, but it also takes more time to do the calculations.

\subsubsection{Directly solving the differential equation}

Probably the simplest way to determine the bound states of a
nonrelativistic system is to directly solve the differential equation.
We have done this by reducing the Schr\"odinger equation to a set of 
coupled first order differential equations and solve them using the
Runge-Kutta method \cite{numrec}. This method is only simple to
implement for a nonrelativistic system. For a relativistic system
it becomes much more difficult or even impossible to calculate the
bound states, since we have to deal with a term that contains a
square root of the differential operator.

\subsubsection{Discretized integral equation}

The radial part of the Schr\"odinger equation in spherical coordinates
is given by
\begin{eqnarray}
 \left(-\frac{\hbar^{2}}{m} \frac{d^2}{dr^2} +
 \frac{\hbar^{2}l(l+1)}{mr^2} + V(r)\right) \psi_{l}(r) =
 E \psi_{l}(r),
\label{radschr}
\end{eqnarray}
where $l$ is the orbital angular momentum quantum number and
$\psi_{l}(r)$ is the radial part of the total wave function which is
defined as
\begin{eqnarray}
 \Psi(\vec{r}) = \psi_{l}(r) Y_{lm}(\hat{r}) =
 \frac{u_{l}(r)}{r} Y_{lm}(\hat{r}) .
\end{eqnarray}
We can also write Eq.~(\ref{radschr}) in the momentum space
representation. We then find the integral equation
\begin{equation}
 \frac{p^{2}}{m}\tilde{\psi}_{l}(p) +
 \int_{0}^{\infty}\mathrm{d}q q^2 \tilde{V}_{l}(p,q)\tilde{\psi}_{l}(q) =
 E\tilde{\psi}_{l}(p).
\end{equation}
Here $\tilde{\psi}_{l}(p)$ is the Fourier transform of the function
$\psi_{l}(r)$ and 
$\tilde{V}_{l}(p,q)$ is the Fourier transformed potential.

Since this integral equation cannot be solved in closed form, we apply
a suitable  Gaussian integration rule, with abscissas $q_{j}$ and weigths
$w_{j}$ and approximate the integral by a finite sum.
We first transform the integration variable $q$ to a variable $x$,
so that the integration interval changes from $[0,\infty)$ to $[-1,1)$.
This is done by the transformation
\begin{equation}
 q(x) = C \, \frac{1 + x}{1 - x},
\end{equation}
where C is a positive scaling parameter. Because of this
transformation, the integrand contains an additional factor, the
Jacobian $\mathrm{d}q(x)/\mathrm{d}x$.

Now that we have approximated the integral by this finite sum, we can
take the momentum variable $p$ equal to the abscissas.  By doing this
we obtain an eigenvalue equation in matrix form, which has the
following form when we choose a total of $N$ abscissas:
\begin{equation}
 \sum_{j=1}^{N}K_{ij}\tilde{\psi}_{l}(q_{j}) = E\tilde{\psi}_{l}(q_{i}),
\label{eq:five}
\end{equation}
where the matrix-elements $K_{ij}$ of the matrix $K$ are given by:
\begin{equation}
 K_{ij} = \frac{q_{j}^{2}}{m}\delta _{ij} +
 \tilde{w}_{j}q^2_j \tilde{V}_{l}(q_{i},q_{j})
\end{equation}
with $\tilde{w} _{j} = (dq(x_{j})/dx_{j}) w_{j}$. 

The matrix equation
(\ref{eq:five}) would be easier to solve if the matrix would be
symmetric. Left-multiplying the matrix equation with a diagonal matrix
$D$, with elements $D_{ij} = q_{j}\delta_{ij}/ \sqrt{\tilde{w}_{j}}$, and
inserting the identity $D^{-1}D$ between the matrix $K$ and the column
vector $\tilde{\psi}_{l}$ gives us a symmetric matrix $\tilde{K} =
DKD^{-1}$. The matrix elements $\tilde{K}_{ij}$ are given by
\begin{equation}
 \tilde{K} _{ij} = \frac{q_{j}^{2}}{m}\delta_{ij} +
 \sqrt{\tilde{w}_{i}\tilde{w}_{j}}v_{l}(q_{i},q_{j})
\end{equation}
for the nonrelativistic situation, and
\begin{equation}
 \tilde{K} {ij} = 2\sqrt{q_{j}^{2} + m^{2}}\delta_{ij} +
 \sqrt{\tilde{w}_{i}\tilde{w}_{j}}v_{l}(q_{i},q_{j})
\end{equation}
for the relativistic situation, with $v_{l}(p,q) = p q \tilde{V}_{l}(p,q)$.
The matrix equation we solve in this approach is given by
\begin{equation}
 \sum_{j=1}^{N}\tilde{K}_{ij}\tilde{u}_{l}(q_{j}) = E\tilde{u}_{l}(q_{i}),
\end{equation}
where $\tilde{u}_{l}$ is the eigenvector, whose $j$-th element is given
by $\sqrt{\tilde{w}_{j}} q_{j}\tilde{\psi}_{l}(q_{j})$.  

\subsubsection{Expanding the eigenfunctions in terms of basis functions}

\begin{figure}[ht]
\begin{center}
 \epsfig{figure=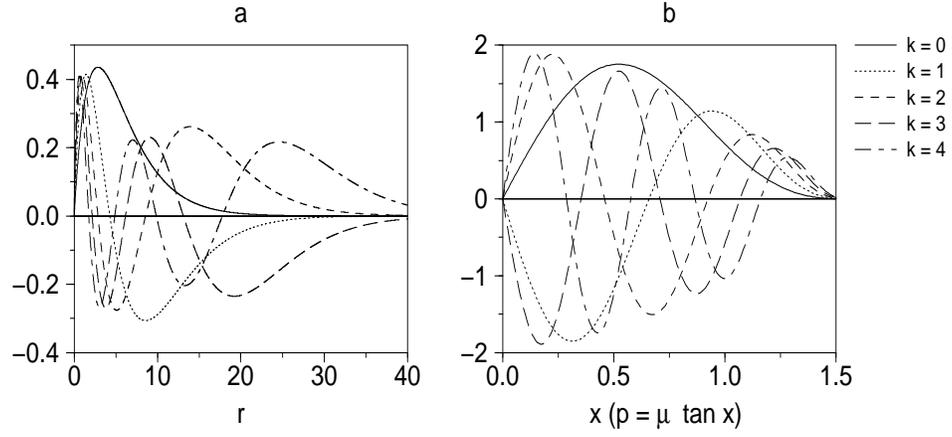,height=6cm,width=12cm,angle=0}
 \caption{First five basis function in configuration space (a)
 and momentum space (b).}
 \label{basefunc}
\end{center}
\end{figure}

In the third approach the eigenvalues are calculated through expanding
the eigenfunctions in terms of basis functions. The basis we use in
this approach is the one first introduced by Olsson \cite{ols} and
Weniger \cite{wen}. The basis functions are known in this basis in both
configuration and momentum space. In configuration space they are given
by
\begin{equation}
\psi_{k,l}(r) = N_{kl} (2\mu r)^{l} e^{-\mu r} L_{k}^{(2l+2)}(2\mu r) \, ,
\end{equation}
where the normalization is given by
\begin{equation}
N_{kl} = \sqrt{\frac{k!(2\mu)^3}{(k + 2l + 2)!}}
\end{equation}
and $L_{k}^{(2l+2)}(2\mu r)$ is an associated Laguerre polynomial.  In
the momentum space representation the basis functions are given by
\begin{equation}
\tilde{\psi}_{k,l}(p) = \tilde{N}_{kl} \,p^{l} \,
\left(\frac{\mu} {p^{2}+\mu^{2}}\right)^{l+2} P_{k}^{(l+3/2,l+1/2)}
\left(\frac{p^{2} - \mu^{2}}{p^{2} + \mu^{2}}\right) \, ,
\end{equation}
with $P^{(\alpha,\beta)}_{k}(x)$ a Jacobi polynomial and the
normalization constant is
\begin{equation}
 \tilde{N}_{kl} = \frac{2\sqrt{\mu k!(k+2l+2)!}}{\Gamma (k+l+3/2)}.
\end{equation}
The first five wave functions are plotted in Fig.~\ref{basefunc} in
both configuration and momentum space.\\ Note that these functions
differ from the familiar Coulomb and harmonic oscillator basis
functions. Ref.~\cite{wen} gives the mathematical details including
the proof of completeness of the basis.\\


With these basis functions the matrix element of the potential between
the initial and final states is calculated in configuration space. As
an example we will calculate the matrix elements for the Yukawa
potential. Other potentials are treated in a similar way. The
angular part of this matrix element reduces to the product of two
Kronecker delta functions. The more interesting part for us is the part
that depends on the radial coordinates; which looks like
\begin{eqnarray}
\int^{\infty}_{0}\!\!\!\mathrm{d}r \,\, u^{*}_{k_{i}l_{i}}(r)
\frac{\exp(-\mu_{\mathrm{Y}}r)}{r} u_{k_{j}l_{j}}(r) \, .
\label{rad}
\end{eqnarray}
The integrand can be rewritten in a somewhat different form. This is
done by first making the following substitution for the reduced radial
wave function
\begin{equation}
u_{kl}(r) = N_{kl} L^{(2l+2)}_{k}(2\mu r) r^{l+1} e^{-\mu r} = \bar{u}_{kl}(r)
e^{-\mu r},
\end{equation}
where $\bar{u}_{kl}(r)$ is a polynomial.  Note that the (reduced)
radial wave functions behave like $r^{l+1}$.  When we make this
substitution, we find for Eq.~(\ref{rad})
\begin{equation}
\int^{\infty}_{0} \!\!\! \mathrm{d}r \frac{1}{r} \exp(-(2\mu + \mu_{
\mathrm{Y}})r) \,\,\bar{u}^{*}_{k_{i}l_{i}}(r)\,\, \bar{u}_{k_{j}l_{j}}(r).
\label{radpart}
\end{equation}
An integral of the form
\begin{equation}
\int^{\infty}_{0}\!\!\!\mathrm{d}x \,\, e^{-x} f(x) ,\nonumber
\end{equation}
where $f(x)$ is a polynomial, can be estimated to the order of machine
precision using Gauss-Laguerre integration. The expression in
Eq.~(\ref{radpart}) can be written in the desired form by first making
the substitution $x = (2\mu + \mu_{Y})r$. Discretizing the equation and
then making another substitution $r_{j} = x_{j}/(2\mu + \mu_{Y})$ gives
the desired form:
\begin{equation}
\sum_{k} \bar{w}_{k} \frac{1}{r_{k}} \bar{u}^{*}_{k_{i}l_{i}}(r_{k}) \,\,
\bar{u}_{k_{j}l_{j}}(r_{k}),
\label{potpart}
\end{equation}
where $\bar{w}_{k}$ is given by
\begin{equation}
\bar{w}_{k} = \frac{w_{k}}{2\mu + \mu_{\mathrm{Y}}} .
\end{equation}
The abscissas $x_k$ and weights $w_k$ are obtained according to the
Gauss-Laguerre rule.

To calculate the bound states of the Hamiltonian given by
Eq.~(\ref{hamrel}) (or its nonrelativistic equivalent), we
expand the eigenfunctions in terms of the basis functions,
$\psi = \sum_{i} a_{i}\psi_{i}$, of the Olsson basis and rewrite
the eigenvalue equation as
\begin{equation}
 \sum_{j} H_{ij} a_{j} =
 \sum_{j}\left< \psi_{i} |H|\psi_{j}\right> a_{j} = E a_{i} .
\label{mateqn}
\end{equation}

\subsection{Determining the accuracy of our methods}

There are several factors that influence the accuracy of our results.
First we should make a distinction between the accuracy obtained
through making a mathematical approximation and the accuracy obtained
because of numerical reasons. The integrals we need when calculating
the matrix elements of the Yukawa potential can be calculated exactly.
However, all other integrals are estimated with a mathematical approximation.

Of course the accuracy of our results depends on more than only the
accuracy with which we can calculate the integrals. We have looked at
the factors that have an influence on the accuracy when we use our
second method (discretizing the integral equation). Here we have a
`scaling' parameter $C$, which has an influence on the accuracy.
Another factor that influences the accuracy is the number of abscissas
we take in our calculation.

In table~\ref{abscis} we give the calculated bound states for one case:
$m = 1.0$, $\rho = -2.0$, $\mu_{\mathrm{Y}} = 0.02$ and a scaling
parameter $C = 1.1$. We have calculated the bound states for five
different numbers of abscissas. From table~\ref{abscis} we see that the
bound states converge.

\begin{table}[ht]
\begin{center}
\caption{Bound state energies for the Yukawa potential with
$m = 1.0$, $\rho = -2.0$ and $\mu_{\mathrm{Y}} = 0.02$ calculated
with different numbers of abscissas.
\label{abscis}}

\vspace{1ex}

\begin{tabular}{|c|c|c|c|c|c|}
\hline
k & 20 absc. & 50 absc. & 80 absc. & 125 absc. & 155 absc. \\
\hline
0 & -1.189770 & -0.998801 & -0.973828 & -0.964817 & -0.962900 \\
1 & -0.279598 & -0.221291 & -0.215165 & -0.213154 & -0.212748 \\
2 & -0.111285 & -0.079688 & -0.077067 & -0.076322 & -0.076185 \\
3 & -0.053701 & -0.032550 & -0.031212 & -0.030875 & -0.030818 \\
4 & -0.028074 & -0.012991 & -0.012274 & -0.012113 & -0.012086 \\
5 & -0.020798 & -0.004369 & -0.004007 & -0.003934 & -0.003923 \\
\hline
\end{tabular}
\end{center}
\end{table}

To see what kind of influence the scaling parameter $C$ has on the
bound states, we have calculated the bound states at different $C$ for
the case $m = 1.0$, $\rho = -2.0$ and $\mu_{\mathrm{Y}} = 0.02$ and
with 155 abcissas. When we look at the results we see that there is a
small domain of scaling parameters where the bound states are stable.
For scaling parameters that lie outside this stable domain, the
calculated bound state energies are larger. In all our calculations
using our second method (discretizing the integral equation) we have
taken a scaling parameter of $C = 1.1$, which lies in the middle of
the stable domain we found.\\

In a similar way we have looked at the factors that have an influence
on the accuracy when we use our third method (expanding the
eigenfunctions in terms of basis functions). For this method the number
of basis functions we have taken into account has an influence on the
accuracy. The parameter $\mu$ is completely free and can be used to
tune the basis to the Hamiltonian at hand. Although this `scaling'
parameter can thus be considered as a variational one, in practice
one should be careful if the integrals one needs to calculate are
obtained numerically. Then it turns out that only in a limited range
of values $\mu$ is of genuinely variational character.

There are several potentials that we have used and for all these
potentials we have looked what kind of influence the variables
mentioned above have on the results. In this report we will only
show it for the Yukawa potential.

\begin{figure}[ht]
\begin{center}
\epsfig{figure=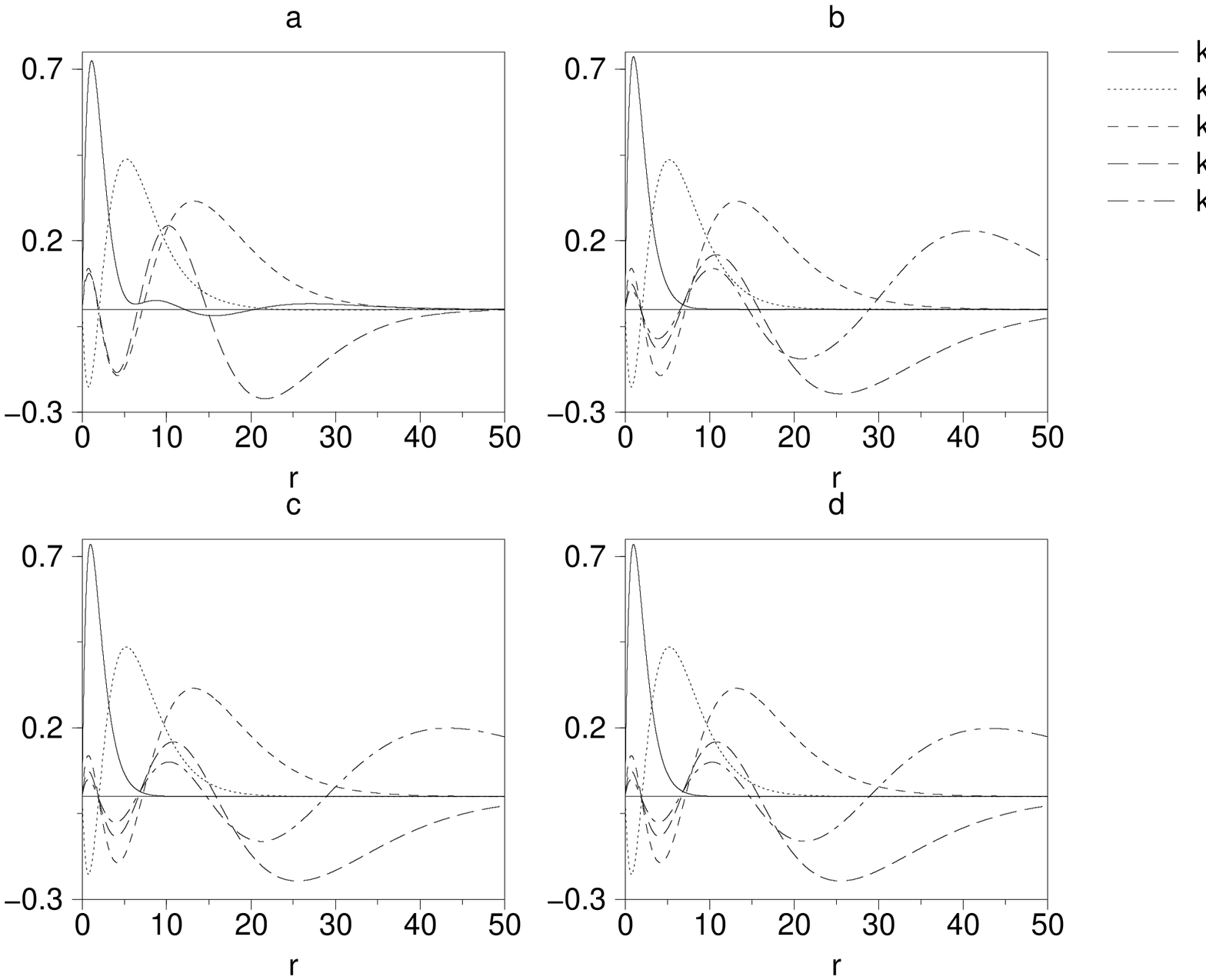,height=10cm,width=10cm,angle=0}
\caption{Wave functions in configuration space calculated with (a) 5,
(b) 10, (c) 15 or (d) 20 basis functions for the Yukawa potential in
the case $m = 1.0$, $\rho = -2.0$ and $\mu_{\mathrm{Y}} = 0.02$.}
\label{rbasfunc}
\end{center}
\end{figure}

\begin{figure}[ht]
\begin{center}
\epsfig{figure=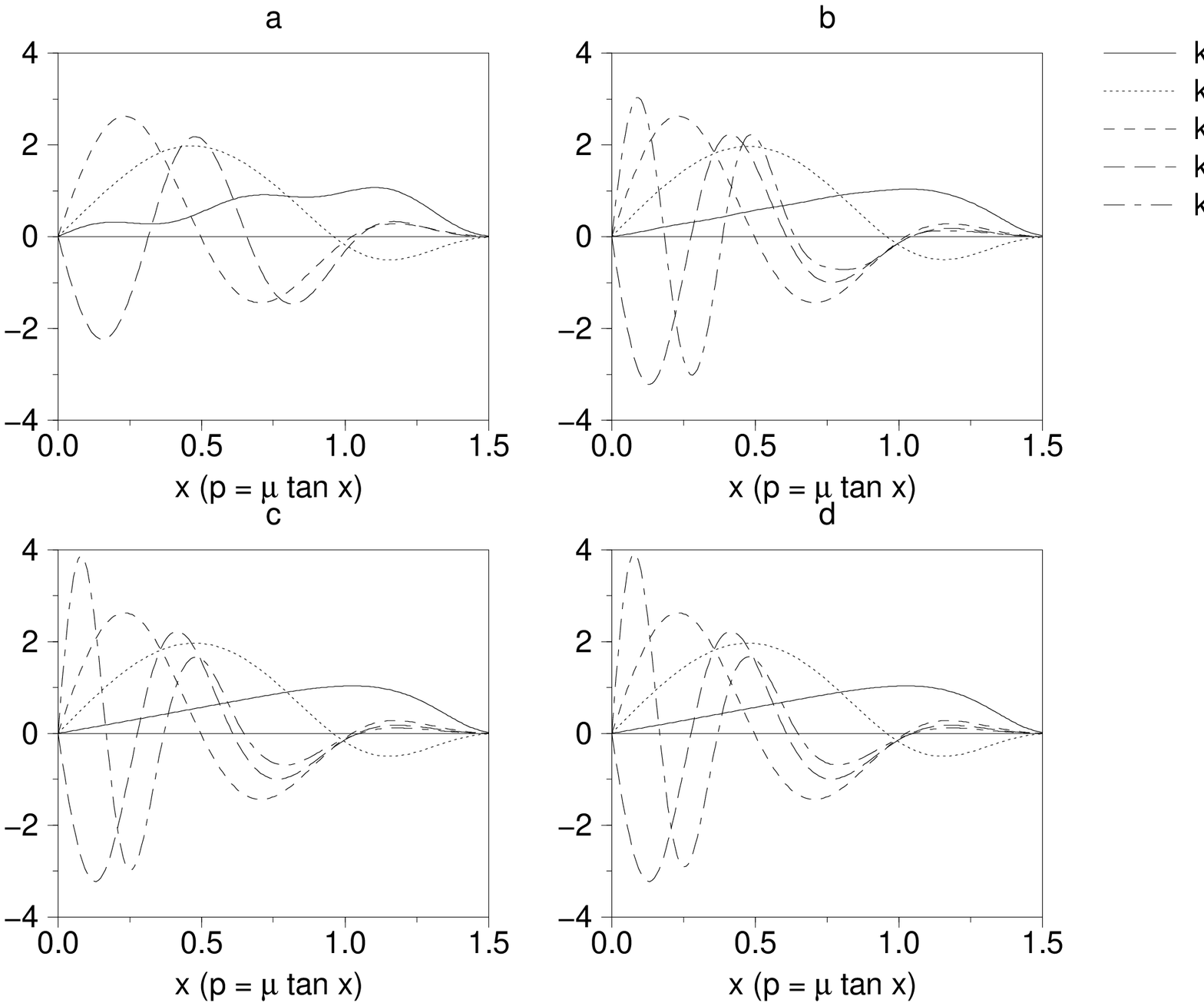,height=10cm,width=10cm,angle=0}
\caption{Wave functions in momentum space calculated with (a) 5,
(b) 10, (c) 15 or (d) 20 basis functions for the Yukawa potential
in the case $m = 1.0$, $\rho = -2.0$ and $\mu_{\mathrm{Y}} = 0.02$.}
\label{pbasfunc}
\end{center}
\end{figure}

We have looked at the influence of the number of basis functions we
take into account. This was done for the following choice of
parameters, which are given below:
\begin{eqnarray}
m = 1.0,  \rho = - 2.0, \mu_{\mathrm{Y}} = 0.02, \mu = 0.35 \nonumber
\end{eqnarray}
We have made an expansion of the eigenfunctions in terms of the first
5, 10, 15 or 20 basis functions; these eigenvalues are given in
table~{\ref{eigvalbasf}}. We have also calculated the wave functions in
both configuration and momentum space, these are given in resp.
Figs.~\ref{rbasfunc} and \ref{pbasfunc}.

\begin{table}[h]
\begin{center}
\caption{Bound state energies for the Yukawa potential with
$m = 1.0$, $\rho = -2.0$ and $\mu_{\mathrm{Y}} = 0.02$ calculated
with different numbers of basis functions.
 \label{eigvalbasf}}

\vspace{1ex}

\begin{tabular}{|c||c|c|c|c|}
\hline
k & 5 basis func. & 10 basis func. & 15 basis func. & 20 basis func. \\
\hline
0 & -0.9419469833 & -0.9605381489 & -0.9605921507 & -0.9605921507 \\
1 & -0.2122372389 & -0.2122966051 & -0.2122966051 & -0.2122966051 \\
2 & -0.07603943348 & -0.07604002953 & -0.07604002953 & -0.07604002953 \\
3 & -0.02350080013 & -0.03075730801 & -0.03075850010 & -0.03075850010 \\
4 & -- & -0.01097238064 & -0.01205313206 & -0.01206004620 \\
5 & -- & -- & -0.002824902534 & -0.003842115402 \\
\hline
\end{tabular}
\end{center}
\end{table}



Looking at the values of the bound state energies given in
table~\ref{eigvalbasf}, we see that these converge. Figs.~\ref{rbasfunc}
and \ref{pbasfunc} show that the wave functions are smoother when
we take more basis functions into account. From this evidence we
conclude that our results can not be taken to be reliable when
only a small number of basis functions is used in making the
expansion. When comparing the results of the expansion to the other
methods we used 20 basis functions.

The `scaling parameter' $\mu$ appears in the basis, but it is
not a real variational parameter since we cannot calculate the matrix
elements of the kinetic energy without making mathematical approximations.
For several values of $\mu_{\mathrm{Y}}$ in the range between 0 and 1
we have looked at the behaviour of $\mu$. When we look at the values
of the ground state energy at one value of $\mu_{\mathrm{Y}}$ and
different $\mu$, we find the same value (for at least the first six
decimal places) of this ground state for $\mu > 0.30$. For $\mu < 0.30$,
we find that the ground state is shifted upwards and has the `wrong'
value. We also have determined a stable domain in $\mu$ where the
first five eigenvalues are the same up to at least six decimal places.
For all values of $\mu_{\mathrm{Y}}$ the stable domain starts at
$\mu = 0.30$ and ends at some value say $\mu_{\mathrm{max}}$. These
maximum values are given in table~\ref{mu} for several values of
$\mu_{\mathrm{Y}}$.

\begin{table}[h]
\begin{center}
\caption{Maximum value of $\mu$ for a stable domain at different
values of $\mu_{\mathrm{Y}}$. \label{mu}}

\vspace{1ex}

\begin{tabular}{|l||l|l|l|l|l|l|l|l|l|l|l|}
\hline
$\mu_{\mathrm{Y}}$ & 0.0 & 0.001 & 0.002 & 0.005 & 0.01 & 0.02 &
0.05 & 0.1 & 0.2 & 0.5 & 1.0\\
\hline
$\mu_{\mathrm{max}}$ & 0.414 & 0.545 & 0.520 & 0.542 & 0.476 &
0.367 & 0.467 & 0.663 & 1.0 & 1.0 & 1.0\\
\hline
\end{tabular}
\end{center}
\end{table}

In most calculations we have done, we have taken a value of $\mu =
0.35$ for the `scaling parameter', which lies in the stable domain
we have determined above.\\

All together we see that both methods (discretizing the integral equation
and making an expansion in basis functions) are not as easy to implement
as the method of directly solving the differential equation, although
it is not extremely difficult either. Both methods have a stable
domain in which we find accurate solutions. These domains are different
for different potentials. The accuracy of both methods is better than
the accuracy reached with directly solving the differential equation.
Comparing the two methods, we see that the results found by making an
expansion in basis functions are more accurate.

\section{Check of the numerical results in the nonrelativistic case}

We have first calculated the bound state energies and wave functions
in the nonrelativistic situation for some potentials for which
we know the solutions. After this we have calculated the bound states
of the Yukawa potential in the nonrelativistic case.

\subsection{The Coulomb potential}

We have calculated the eigenvalues for the Coulomb potential only
with the third method. The reason is that the long-range character
of the Coulomb potential leads to a singularity in momentum space
that is difficult to handle numerically and in configuration space
to an uncertainty whether the asymptotic region is reached in the
differential equation.

We limited ourselves to making an expansion of the eigenfunctions
in 20 basis states and checked that the first five eigenvalues
are accurate. In the calculation we have taken the mass $m = 1.0$,
the strength of the potential $\rho = -2.0$ and the `scaling parameter'
$\mu = 0.35$.

The calculated eigenvalues and the exact eigenvalues of the Coulomb
potential are given in table~\ref{coul} for the first eight eigenvalues.

\begin{table}[h]
\begin{center}
\caption{The calculated eigenvalues ($l = 0$) of the Yukawa potential with
 $\mu_{\mathrm{Y}} = 0$ and the exact eigenvalues of the Coulomb potential.
 \label{coul}}

\vspace{1ex}

\begin{tabular}{|c|c|c|}
\hline
n & exact & calculated \\
\hline
1 & $-1.0000000000$ & $-1.0000000000$ \\
2 & $-0.2500000000$ & $-0.2500000000$ \\
3 & $-0.1111111111$ & $-0.1111111641$ \\
4 & $-0.0625000000$ & $-0.0625000000$ \\
5 & $-0.0400000000$ & $-0.0399999619$ \\
6 & $-0.0277777777$ & $-0.0277774334$ \\
7 & $-0.0204081630$ & $-0.0202956199$ \\
8 & $-0.0156250000$ & $-0.0137485265$ \\
\hline
\end{tabular}
\end{center}
\end{table}

\subsection{The linear potential}

We have calculated the bound states for the linear potential only with
the method of expanding the eigenfunctions in terms of the Olsson basis
functions. The bound states are calculated in the nonrelativistic
situation for $m = 1.0$, $\sigma = 1.0$, $\mu = 0.90900$ 
and are given in table~\ref{airyzero}.

Since $\sigma = 1.0$ and $m = 1.0$ the bound states should be equal to
the zeros of the Airy function. Comparing the results in table~\ref{airyzero}
with the zeros of the Airy function, we see that the first excited state
is accurate to the seventh decimal. Higher excited states are less
accurate; the accuracy decreases to an accuracy of $\approx 1:8000$ for
the fourth excited state.

\begin{table}[ht]
\begin{center}
\caption{Zeros of the Airy function and eigenvalues for the nonrelativistic
 Schr\"odinger equation with $m=1$, $\sigma =1$. The parameter $\mu$ in
the basis has the value $\mu = 0.909$. \label{airyzero}}

\vspace{1ex}

\begin{tabular}{|c|c|c|}
\hline
n & Airy zero & eigenvalue\\
\hline
1 & 2.33810741 & 2.33810741 \\
2 & 4.08794944 & 4.08794976 \\
3 & 5.52055983 & 5.52057237 \\
4 & 6.78670809 & 6.78687212 \\
5 & 7.94413359 & 7.94534443 \\
\hline
\end{tabular}
\end{center}
\end{table}

\begin{figure}[ht]
\begin{center}
\epsfig{figure=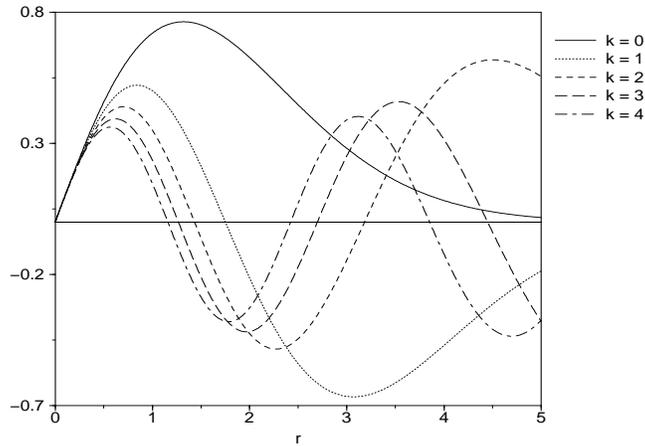,height=6cm,width=8cm,angle=0}
 \caption{Configuration space wave functions obtained by making an expansion
 in basis functions. Linear potential with $m = 1.0$, $\sigma = 1.0$.
 \label{linwave}}
\end{center}
\end{figure}

In this case we have also calculated the wave functions. These are
plotted in Fig.~\ref{linwave}.


\subsection{The Hulth\'en potential}

For the Hulth\'en potential we have calculated the bound state energies
with our third method. In table~\ref{hul} the calculated and exact
bound states are given for the case $m = 1.0$, $\rho = -0.5$ and
$\mu_{\mathrm{H}} = 0.1$. As `scaling' parameter we have used here
$\mu = 1.0$ and we have made an expansion in terms of the first 20
basis states.

\begin{table}[h]
\begin{center}
\caption{Calculated and exact bound state energies for the Hulth\'en
potential for $m = 1.0$, $\rho = -0.5$ and $\mu_{\mathrm{H}} = 0.1$.
 \label{hul}}
 \vspace{1ex}

\begin{tabular}{|c|c|c|}
\hline
k & calculated & exact\\
\hline
0 & -6.002500057 & -6.002500000 \\
1 & -1.322499990 & -1.322500000 \\
2 & -0.466944456 & -0.466944444 \\
3 & -0.180624962 & -0.180625000 \\
4 & -0.062495589 & -0.062500000 \\
5 & -0.010959864 & -0.013611111 \\
\hline
\end{tabular}
\end{center}
\end{table}

We see that the first five calculated and exact bound states agree to
at least four decimal places. Only the very weakly bound fifth excited
state cannot be determined accurately.

\begin{figure}[ht]
\begin{center}
 \epsfig{figure=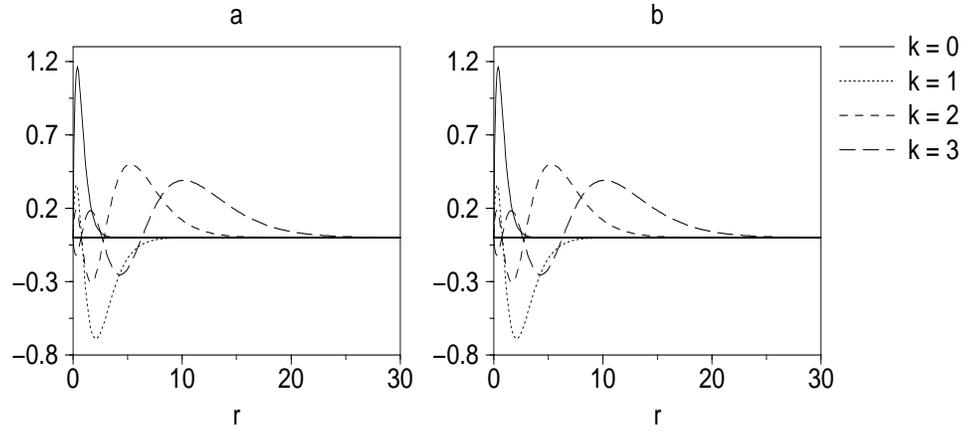,height=6cm,width=12cm,angle=0}
 \caption{Calculated wave functions (by making an expansion in basis
 functions) (a) and exact wave functions (b) in configuration space for
 the Hulth\'en potential in the case: $m = 1.0$, $\rho = -0.5$ and
 $\mu_{\mathrm{H}} = 0.1$.}
 \label{hulwave1}
\end{center}
\end{figure}

\begin{figure}[ht]
\begin{center}
 \epsfig{figure=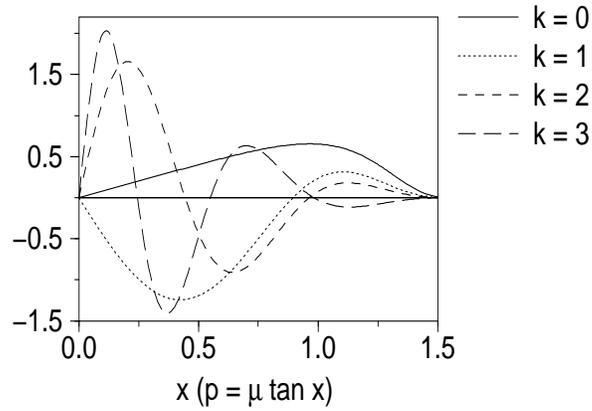,height=6cm,width=8cm,angle=0}
 \caption{Calculated wave functions in momentum space for the Hulth\'en
 potential for the case: $m = 1.0$, $\rho = -0.5$ and
 $\mu_{\mathrm{H}} = 0.1$.}
 \label{hulwave2}
\end{center}
\end{figure}

In Fig.~\ref{hulwave1} the exact wave functions and the
wave functions calculated by making an expansion into 20 basis states
are plotted in configuration space. We see that the wave functions we
have calculated are the same as the exact wave functions. In
Fig.~\ref{hulwave2} the calculated wave functions are plotted in
momentum space.

%

\begin{figure}[ht]
\begin{center}
 \epsfig{figure=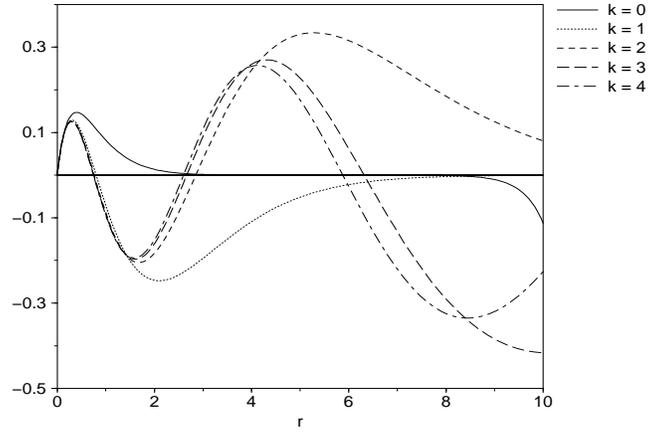,height=6cm,width=8cm,angle=0}
 \caption{Wave functions calculated for the Hulth\'en potential by directly
 solving the differential equation for the case: $m = 1.0$, $\rho = -0.5$
 and $\mu_{\mathrm{H}} = 0.1$.}
 \label{hulwave3}
\end{center}
\end{figure}

We have also calculated the wave functions by directly solving the
differential equation (Fig.~\ref{hulwave3}). Here we have used the
bound state energies calculated with the third approach as an input.


Note that the wave functions in Fig.~\ref{hulwave3} are not normalized.
To see whether these wave functions are the same as the exact and
calculated ones in Fig.~\ref{hulwave1} we can look at the location of
the nodes and of the extrema. In table~\ref{hulzero} the nodes and extrema
are given for the exact wave functions and the wave functions calculated
by directly solving the differential equation.
\begin{table}[h]
\begin{center}
\caption{Nodes and extrema of the exact wave functions and the calculated
wave functions by directly solving the differential equation for the
Hulth\'en potential. \label{hulzero}}

\vspace{1ex}

\begin{tabular}{|c|c|c|c|c|}
\hline
 & \multicolumn{2}{c|}{nodes} & \multicolumn{2}{c|}{extrema} \\
\hline
k & exact & calculated & exact & calculated \\
\hline
0 & -- & -- & 0.400 & 0.400 \\
 & & & & \\
1 & 0.8004 & 0.800 - 0.804 & 0.304 & 0.304\\
 & & & 2.098 & 2.098 \\
 & & & & \\
2 & 0.7616 & 0.760 - 0.764 & 0.296 & 0.296 \\
  & 2.8514 & 2.848 - 2.852 & 1.680 & 1.680  \\
 & & & 5.278 & 5.278 \\
 & & & & \\
3 & 0.7501 & 0.750 - 0.752 & 0.292 & 0.292 \\
  & 2.6632 & 2.660 - 2.664 & 1.608 & 1.608 \\
  & 6.3132 & 6.312 - 6.316 & 4.314 & 4.312 - 4.316 \\
 & & & 10.146 & 10.140 - 10.148 \\
\hline
\end{tabular}
\end{center}
\end{table}
Looking at table~\ref{hulzero} we see that the nodes and the
maxima coincide.

\subsection{The Yukawa potential}

In the case of the Yukawa potential (with $\mu_{\mathrm{Y}} \neq 0$ in
Eq.~(\ref{pot})) there are no exact solutions known. We do know,
however, that the eigenvalues in this case are greater than the
corresponding eigenvalues in the Coulomb case. We also know that the
eigenvalues should vary smoothy when $\mu_{\mathrm{Y}}$ is varied.

To find the stable domain in which we want to calculate the eigenstates,
we have varied $\mu$ for different values of $\mu_{\mathrm{Y}}$. 
In all calculations we have used a value of $\mu = 0.35$, which is 
in the stable domain for all values of $\mu_{\mathrm{Y}}$. 
Now that we know where the stable domain lies, we have calculated
the eigenstates for the Yukawa potential for several values of
$\mu_{\mathrm{Y}}$. In every case we have used 20 basis states in the
expansion and expect the first five states to be accurate. All the
eigenvalues are calculated for a mass $m = 1.0$ and a strength of the
potential $\rho = -2.0$. The results are given in table~\ref{yuk}.

\begin{table}[ht]
\begin{center}
\caption{Eigenvalues ($l = 0$) of the Yukawa potential for different
values of $\mu_{\mathrm{Y}}$ in the nonrelativistic case $m = 1.0$
and $\rho = -2.0$. (a `$--$' means a bound state has not been found)
\label{yuk}}

\vspace{1ex}

\begin{tabular}{|c|c|c|c|c|}
\hline
k & $\mu_{\mathrm{Y}}=0.01$ & $\mu_{\mathrm{Y}}=0.02$ &
$\mu_{\mathrm{Y}}=0.05$ & $\mu_{\mathrm{Y}}=0.1$ \\
\hline
0 & $-0.9801490307$ & $-0.9605921507$ & $-0.9036328793$ & $-0.8141160011$ \\
1 & $-0.2305865288$ & $-0.2122966051$ & $-0.1635423899$ & $-0.0998564959$ \\
2 & $-0.0923976898$ & $-0.0760400295$ & $-0.0387051106$ & $-0.0064160824$ \\
3 & $-0.0447121859$ & $-0.0307585001$ & $-0.0061832666$ & $--$ \\
4 & $-0.0233221054$ & $-0.0120600462$ & $--$ & $--$ \\
\hline
\end{tabular}
\newline
\newline
\newline

\begin{tabular}{|c|c|c|c|}
\hline
k & $\mu_{\mathrm{Y}}=0.2$ & $\mu_{\mathrm{Y}}=0.5$ &
$\mu_{\mathrm{Y}}=1.0$ \\
\hline
0 & $-0.6536170244$ & $-0.2962340117$ & $-0.0205712318$ \\
1 & $-0.0242156982$ & $--$ & $--$ \\
\hline
\end{tabular}
\end{center}
\end{table}

\begin{figure}[ht]
\begin{center}
\epsfig{figure=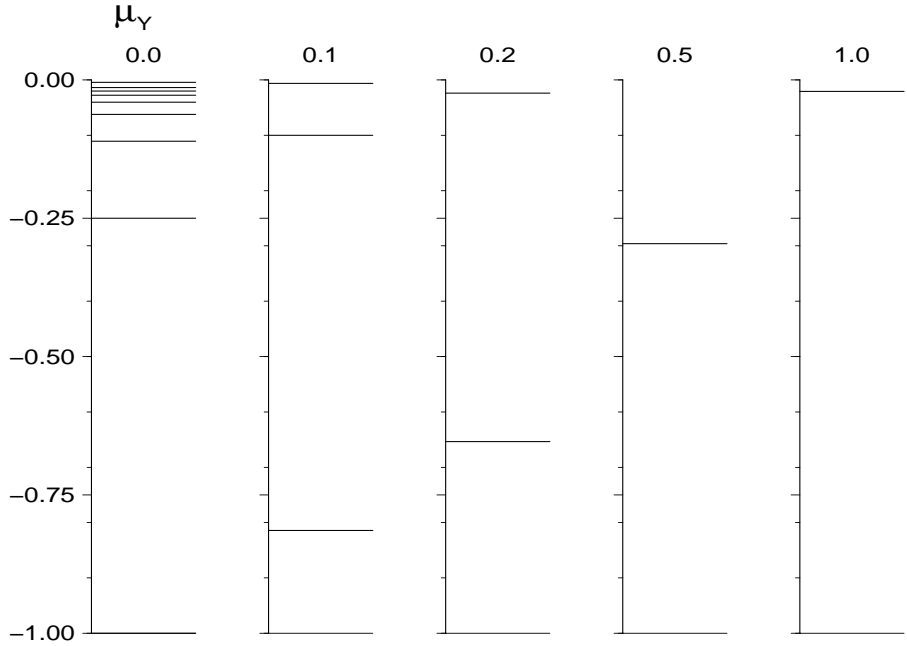,height=8cm,width=12cm,angle=0}
\caption{Bound states of a nonrelativistic $q\bar{q}$ system in a
Yukawa potential for different values of $\mu_{\mathrm{Y}}$.}
\label{specplot}
\end{center}
\end{figure}

We see that for small $\mu_{\mathrm{Y}}$ there are several bound
states. As $\mu_{\mathrm{Y}}$ increases, the number of bound states
decreases. For $\rho = -2.0$ and $\mu_{\mathrm{Y}} > 1.8$ there are no
bound states found at all. The calculated spectra are also plotted in
Fig.~\ref{specplot}. In these plots we see very clearly that all
energies shift upwards, when $\mu_{\mathrm{Y}}$ increases.   

 
We have not only looked at the eigenvalues, but also at the wave
functions. For one case we give here the eigenvalues and wave
functions as we have calculated them with the different methods. The
calculated eigenvalues for the case
\begin{equation}
m = 1.0, \rho = -1.0, \mu_{\mathrm{Y}} = 0.02
\end{equation}
are given in table~\ref{yukval}. Note that only the bound states
calculated with the second and third method are given in this table.
For the first method (directly solving the differential equation) we
have used the bound state energies calculated with our third method 
as an input to calculate the wave functions.

\begin{table}[ht]
\begin{center}
\caption{Eigenvalues (nonrelativistic) for the case $m = 1.0$,
$\rho = -1.0$ and $\mu_{\mathrm{Y}} = 0.02$ calculated by discretizing
the integral equation (second method) and making an expansion of
the eigenfunctions (third method). \label{yukval}}

\vspace{1ex}

\begin{tabular}{|c|c|c|}
\hline
k & integral eqn. & exp. in basis \\
\hline
0 & -0.230696 & -0.2305848598 \\
1 & -0.044726 & -0.0447072983 \\
2 & -0.012351 & -0.0123461485 \\
3 & -0.002981 & -0.0029529333 \\
\hline
\end{tabular}
\end{center}
\end{table}
 
\begin{figure}[ht]
\begin{center}
\epsfig{figure=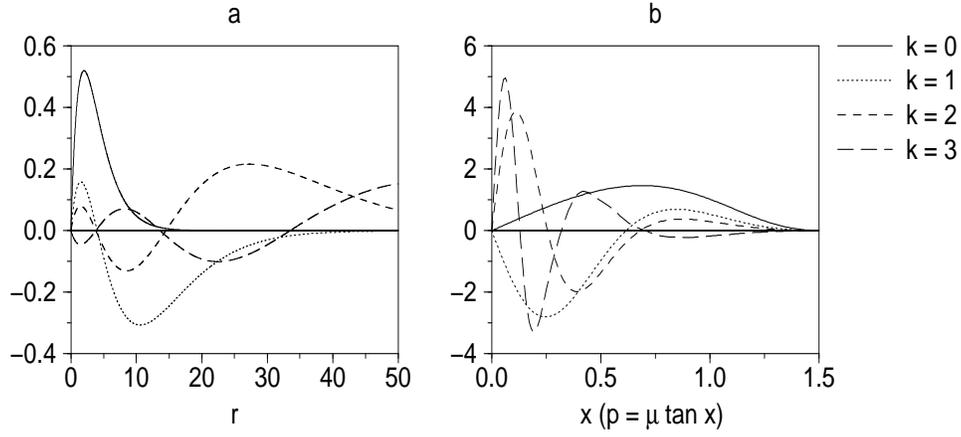,height=6cm,width=12cm,angle=0}
\caption{Wave functions in (a) configuration space and (b) momentum
space calculated for the Yukawa potential by making an expansion of
the eigenfunctions in basis functions for the case: $m = 1.0$, $\rho
= -1.0$ and $\mu_{\mathrm{Y}} = 0.02$.}
\label{yukwave1}
\end{center}
\end{figure}

Using all three methods, we have calculated the wave functions. In
Fig.~\ref{yukwave1} the wave functions in both configuration space and
momentum space are given as they are calculated by expanding the
eigenfunctions in terms of the basis functions.

\begin{figure}[ht]
\begin{center}
\epsfig{figure=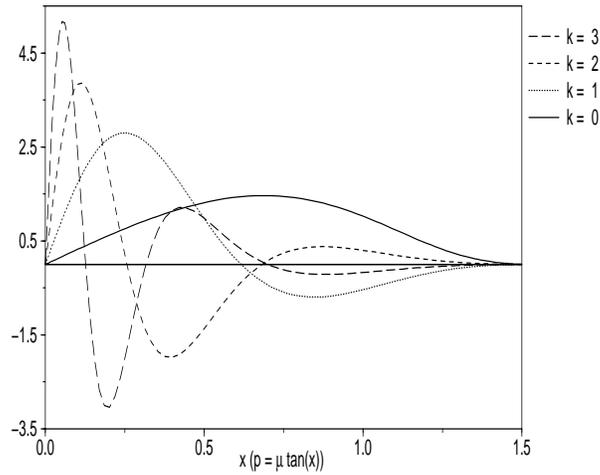,height=8cm,width=6cm,angle=-90}
\caption{Wave functions in momentum space calculated for the Yukawa
potential by directly discretizing the integral equation for the
case: $m = 1.0$, $\rho = -1.0$ and $\mu_{\mathrm{Y}} = 0.02$.}
\label{yukwave2}
\end{center}
\end{figure}

 
We have also calculated the wave function in momentum space by the
method of discretizing the integral equation (Fig.~\ref{yukwave2}).   

\begin{figure}[ht]
\begin{center}
\epsfig{figure=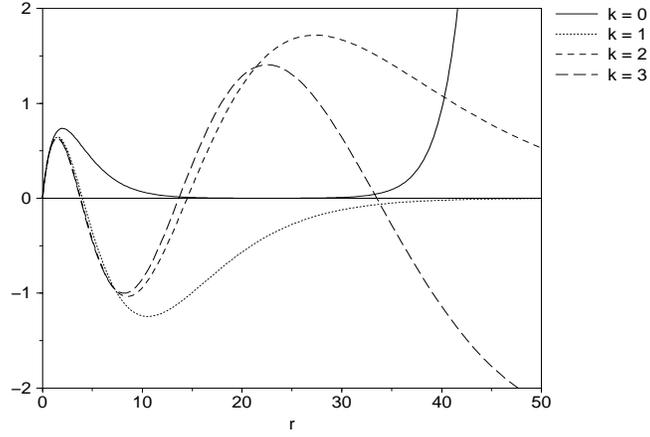,height=6cm,width=8cm,angle=0}
\caption{Wave functions in configuration space calculated for the
Yukawa potential by directly solving the differential equation for
the case: $m = 1.0$, $\rho = -1.0$ and $\mu_{\mathrm{Y}} = 0.02$.}
\label{yukwave3}
\end{center}
\end{figure}

 
Finally we have calculated the wave functions in coordinate space
by directly solving the differential equation. We have taken the
eigenvalues calculated with our third method as input. The wave
functions calculated by this method are plotted in Fig.~\ref{yukwave3}.

%
Looking at this figure, we see that the wave functions at the end show
some strange behaviour, that is they do not converge. The reason that
these wave functions do not converge is due to the accumulation of errors
Note that the wave functions shown in Fig.~\ref{yukwave3} are not
normalized. When the wave functions calculated with both methods are the
same their nodes and extrema must coincide. In table~\ref{zeropoints}
estimates for hte nodes and extrema are given.

\begin{table}[h]
\begin{center}
\caption{Nodes and extrema of the first four wave functions of the
nonrelativistic Yukawa Hamiltonian calculated
with the first method (directly solving the differential equation) and the
third method (expanding the eigenfunctions in terms of basis functions). The
parameters are $m = 1.0$, $\rho = -1.0$ and $\mu_{\mathrm{Y}} = 0.02$.
\label{zeropoints}}

\vspace{1ex}

\begin{tabular}{|c|c|c|c|c|}
\hline
 & \multicolumn{2}{c|}{nodes} & \multicolumn{2}{c|}{extrema} \\
\hline
k & solving diff. eqn. & exp. in basis & solving diff. eqn. & exp. in basis \\
\hline
0 &  -- & -- & 1.98 - 2.01 & 2.00 - 2.01 \\
 & & & & \\
1 & 3.97 - 4.04 & 4.01 - 4.02 & 1.50 - 1.53 & 1.53 - 1.54 \\
 & & & 10.51 - 10.61 & 10.56 - 10.58 \\
 & & & & \\
2 & 3.82 - 3.88 & 3.82 - 3.83 & 1.48 - 1.50 & 1.48 - 1.49 \\
  & 14.45 - 14.59 & 14.50 - 14.51 & 8.44 - 8.52 & 8.49 - 8.50 \\
 & & & 27.28 - 27.50 & 27.28 - 27.33 \\
 & & & & \\
3 & 3.73 - 3.79 & 3.78 - 3.79 & 1.47 - 1.50 & 1.46 - 1.47 \\
  & 13.54 - 13.66 & 13.64 - 13.65 & 8.16 - 8.24 & 8.20 - 8.21 \\
  & 33.45 - 33.72 & 33.46 - 33.47 & 22.46 - 22.62 & 22.41 - 22.45 \\
\hline
\end{tabular}
\end{center}
\end{table}
 
We see that the zero points of the wave functions lie in the same
place, which means that we are dealing with the same wave functions.   

\section{Physical interesting cases}

The physically more interesting situation is the relativistic situation.
For the Yukawa potential we have calculated the bound state energies and
wave functions in the relativistic situation. In this case we see that
the wave functions collapse above a certain strength of the potential.
This collapse occurs only for certain types of potentials.

\subsection{The relativistic case}
 
\begin{figure}[ht]
\begin{center}
\epsfig{figure=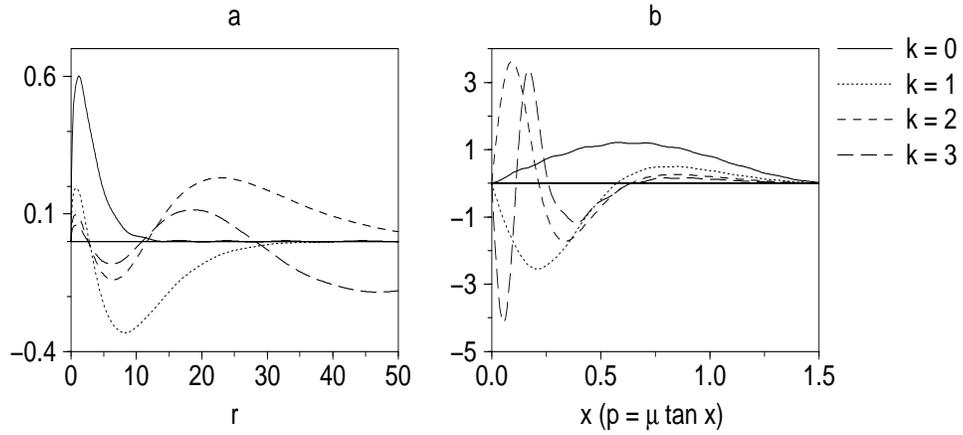,height=6cm,width=12cm,angle=0}
\caption{Wave functions for the Yukawa potential by making an
expansion in basis functions for the case: $m = 1.0$, $\rho = -1.0$
and $\mu_{\mathrm{Y}} = 0.02$ in the relativistic situation.}
\label{yukwave4}
\end{center}
\end{figure}

We have done some calculations in the relativistic case for the
Yukawa potential. In this situation the bound states  are calculated
with the method of expanding the eigenfunctions in terms of basis
functions only. We have looked at the same case as in the nonrelativistic
situation: $m = 1.0$, $\rho = -1.0$ and $\mu_{\mathrm{Y}} = 0.02$.
The results are given in table~\ref{yukrel}. For comparison we have
given the nonrelativistic bound states in the same table.
The wave functions are plotted in Fig.~\ref{yukwave4}.
\begin{table}[ht]
\begin{center}
\caption{Calculated bound state energies (relativistic and
nonrelativistic) for the Yukawa potential in the case $m = 1.0$,
$\rho = -1.0$ and $\mu_{\mathrm{Y}} = 0.02$. \label{yukrel}}

\vspace{1ex}

\begin{tabular}{|c|c|c|}
\hline
k & relativistic & nonrelativistic \\
\hline
0 & -0.3107976913 & -0.2305848598 \\
1 & -0.0583696365 & -0.0447072983 \\
2 & -0.0161626339 & -0.0123461485 \\
3 & -0.0042042733 & -0.0029529333 \\
\hline
\end{tabular}
\end{center}
\end{table}
%
 
\subsection{The collapse of wave functions}
 
It is interesting to look at what happens in the relativistic situation
to the bound states and the wave functions when the strength
($\rho$) of the potential increases. We know that for the Coulomb
potential there exists a critical value of the strength above which the
wave functions collapse \cite{ray, grosmar}. We expect to find this
collapse for the Yukawa potential as well, since this potential is also
not bounded from below.

With a variational calculation we are able to estimate this critical
value for the Yukawa potential. As the variational wave function we
have taken the lowest Olsson wave function (which is the same as the
lowest Coulomb wave funtion). These are given in coordinate space by
\begin{eqnarray}
\psi(r) = \sqrt{4\mu^3} e^{-\mu r}
\end{eqnarray}
and in momentum space by
\begin{eqnarray}
\tilde{\psi}(p) = \sqrt{\frac{32\mu^5}{\pi}} \frac{1}{(p^2 + \mu^2)^2} .
\end{eqnarray}
Using these trial functions we can calculate the potential energy exactly:
\begin{eqnarray}
V(\mu) & = & \int_{0}^{\infty} dr r^2 \sqrt{4\mu^3} e^{-\mu r} \rho
\frac{e^{-\mu_{\mathrm{Y}}r}}{r} \sqrt{4\mu^3} e^{-\mu r} \nonumber\\
& = & \frac{4\mu^3 \rho}{(2\mu + \mu_{\mathrm{Y}})^2} .
\end{eqnarray}
The kinetic energy is given by the integral
\begin{eqnarray}
T(\mu) & = & \int_{0}^{\infty} dp p^2 \sqrt{\frac{32\mu^5}{\pi}}
\frac{1}{(p^2 + \mu^2)^2} 2\sqrt{p^2 + m^2} \sqrt{\frac{32\mu^5}{\pi}}
\frac{1}{(p^2 + \mu^2)^2} \nonumber\\
& = & \frac{64\mu^5}{\pi} \int_{0}^{\infty} dp \frac{p^2 \sqrt{p^2 + m^2}}
{(p^2 + \mu^2)^4} \, .
\label{kinen}
\end{eqnarray}
In the limit $\mu \rightarrow \infty$ we can estimate the kinetic
energy. Defining $\beta = m / \mu$ and $p = \mu x$ and substituting
this in Eq.~(\ref{kinen}), we get
\begin{eqnarray}
T(\mu) & = & \frac{64\mu}{\pi}\int_{0}^{\infty} dx \frac{x^2
\sqrt{x^2 + \beta^2}}{(x^2 + 1)^4}\nonumber\\
& \sim & \frac{64\mu}{\pi}\int_{0}^{\infty} dx \frac{x^3}{(x^2 + 1)^4}
\nonumber\\
& = & \frac{64\mu}{12 \pi} .
\end{eqnarray}
In the second step we have taken the limit $\mu \rightarrow \infty$, or
$\beta \rightarrow 0$. In this limit, the potential energy becomes:
\begin{eqnarray}
V(\mu) = \frac{4\mu^3 \rho}{4 \mu^2} = \rho \mu \, .
\end{eqnarray}
For the total Hamiltonian in this limit we find
\begin{eqnarray}
H(\mu) = \frac{16\mu}{3 \pi} + \rho \mu \, .
\end{eqnarray}
From this Hamiltonian we can determine the critical value, for which
the potential is not bounded from below any more.
\begin{eqnarray}
\rho_{crit} = -\frac{16}{3\pi} \approx -1.697652726
\end{eqnarray}

A more accurate estimation is given by \cite{ray}: $\rho_{crit} =
-4/\pi = -1.2732395$.

\begin{figure}[ht]
\begin{center}
\epsfig{figure=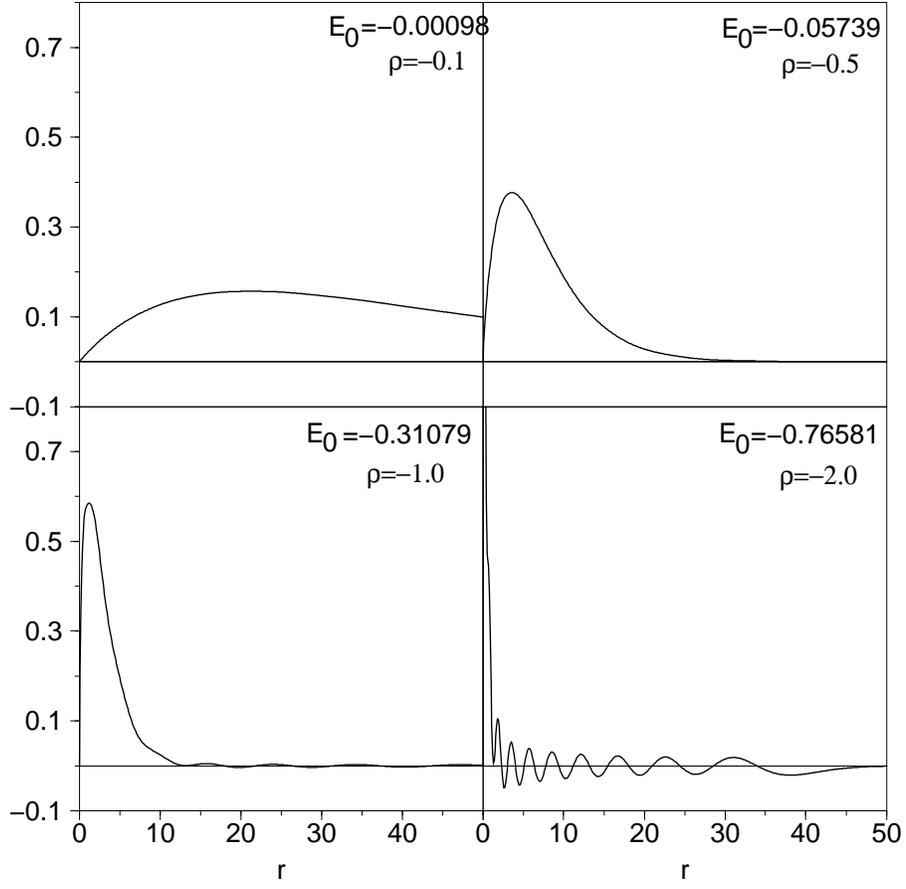,height=12cm,width=12cm,angle=0}
\caption{Wave function (relativistic) of the ground state in
configuration space for the Yukawa potential with $m= 1.0$,
$\mu_{\mathrm{Y}} = 0.02$ and several values of $\rho$.}
\label{yukcolr}
\end{center}
\end{figure}

\begin{figure}[ht]
\begin{center}
\epsfig{figure=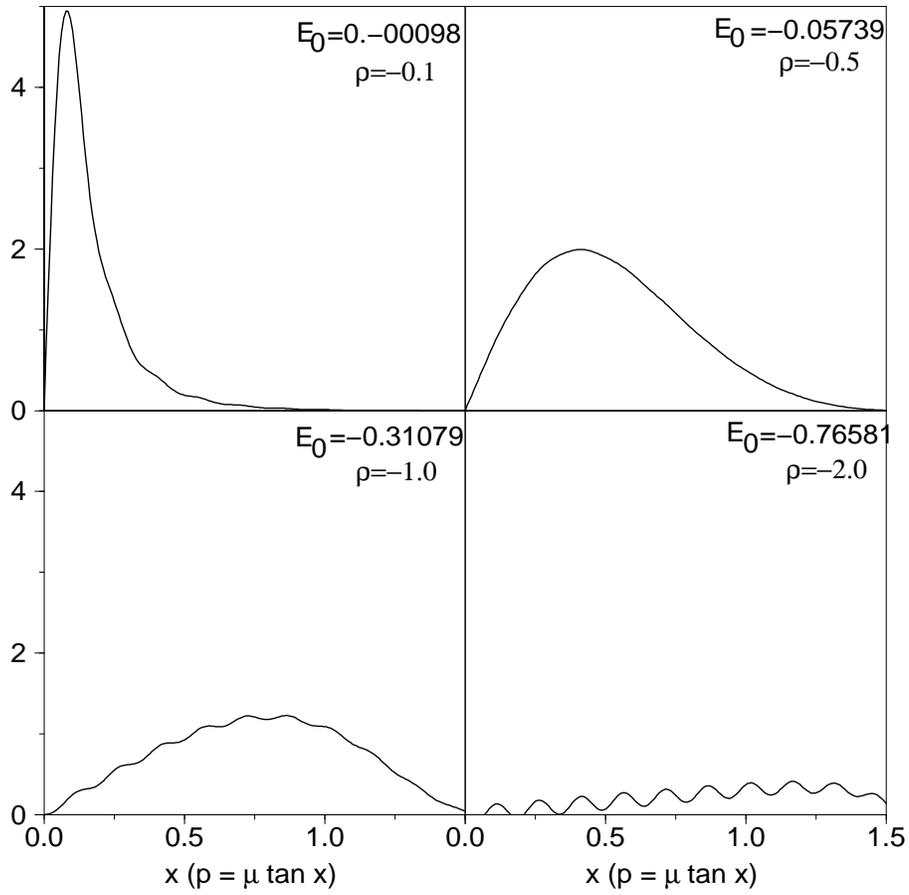,height=12cm,width=12cm,angle=0}
\caption{Wave function (relativistic) of the ground state in
momentum space for the Yukawa potential with $m = 1.0$,
$\mu_{\mathrm{Y}} = 0.02$ and several values of $\rho$.}
\label{yukcolp}
\end{center}
\end{figure}

To see whether this collapse really occurs, we have looked at what
happens when we increase the strength $\rho$ and keep all other
parameters the same.  In Fig.~\ref{yukcolr} we have plotted the wave
functions of the ground state in configuration space for $m = 1.0$,
$\mu_{\mathrm{Y}} = 0.02$ and several values of $\rho$. In
Fig.~\ref{yukcolp} the same is done for the wave functions in the
momentum space representation. In both figures we see that the wave
functions show some irregularities at larger $\rho$. For $\rho \lesssim
\rho_{crit}$ the wave function is not a smooth function anymore and
ceases to be approximated well by our basis. So we get some
irregularities due to numerical errors.


 
\begin{figure}[ht]
\begin{center}
\epsfig{figure=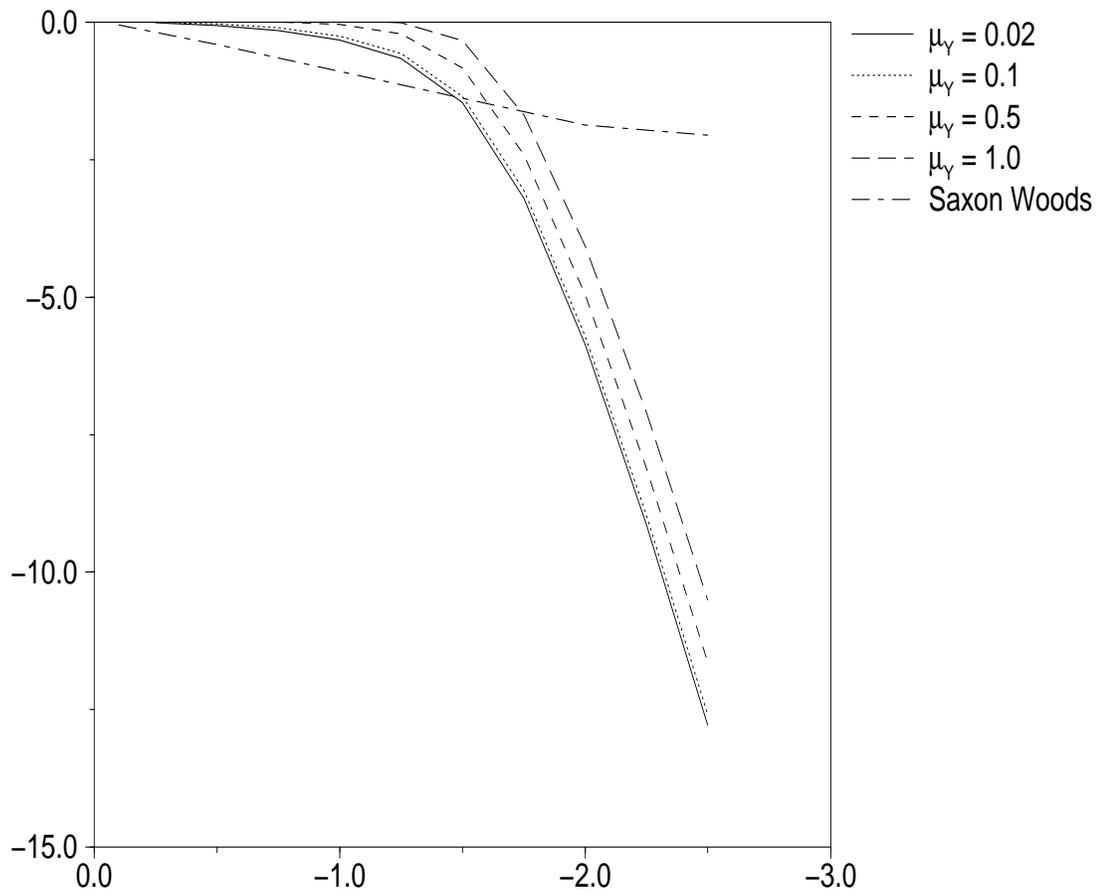,height=12cm,width=12cm,angle=0}
\caption{The strength $\rho$ vs. the ground state energy for several
values of $\mu_{\mathrm{Y}}$.}
\label{rhoenergy}
\end{center}
\end{figure}

To illustrate that the ground state energy drops very fast, we
have plotted it in Fig.~\ref{rhoenergy} for different values of
$\rho$ and several values of $\mu_{\mathrm{Y}}$. We see that the
curves for different $\mu_{\mathrm{Y}}$ do not coincide,
but relative differences are getting smaller for larger $\rho$.

\begin{figure}[ht]
\begin{center}
\epsfig{figure=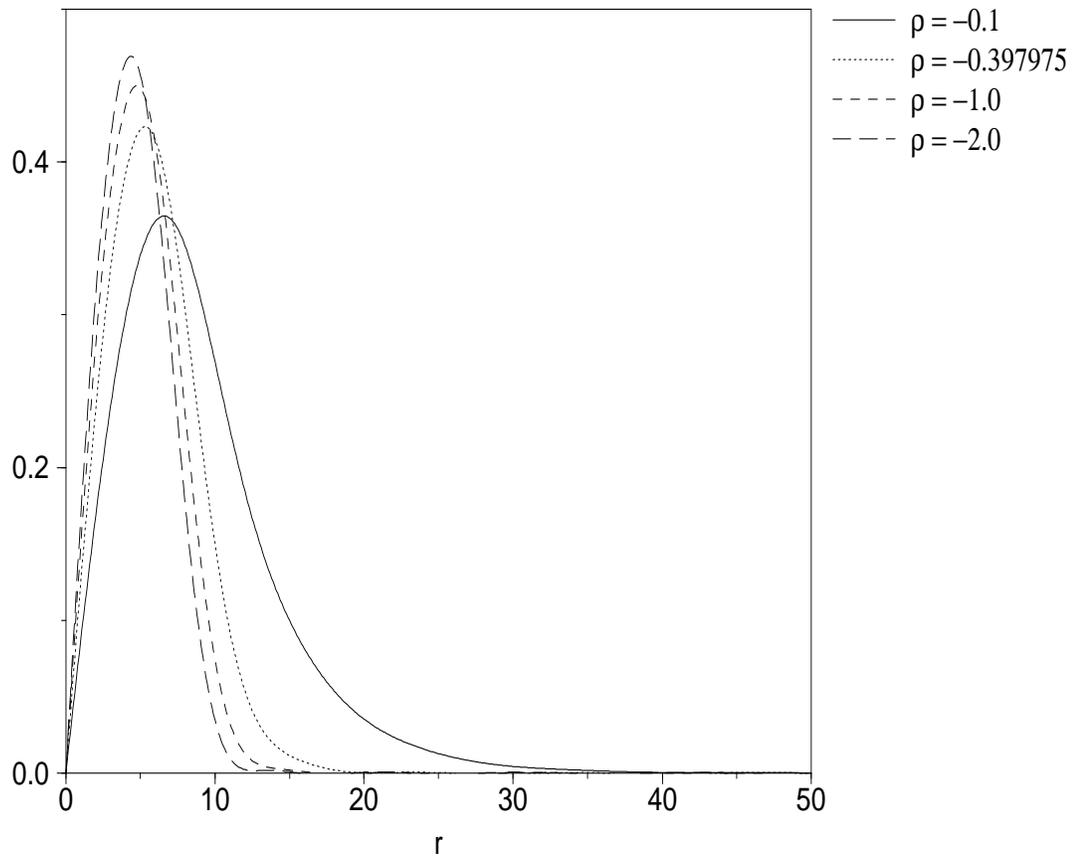,height=12cm,width=12cm,angle=0}
\caption{Wave function (relativistic) of the ground state in
configuration space for the Saxon Woods potential with $m = 1.0$,
$R = 10.0$, $a = 1.0$.}
\label{saxwoodrwv}
\end{center}
\end{figure}

\begin{figure}[ht]
\begin{center}
\epsfig{figure=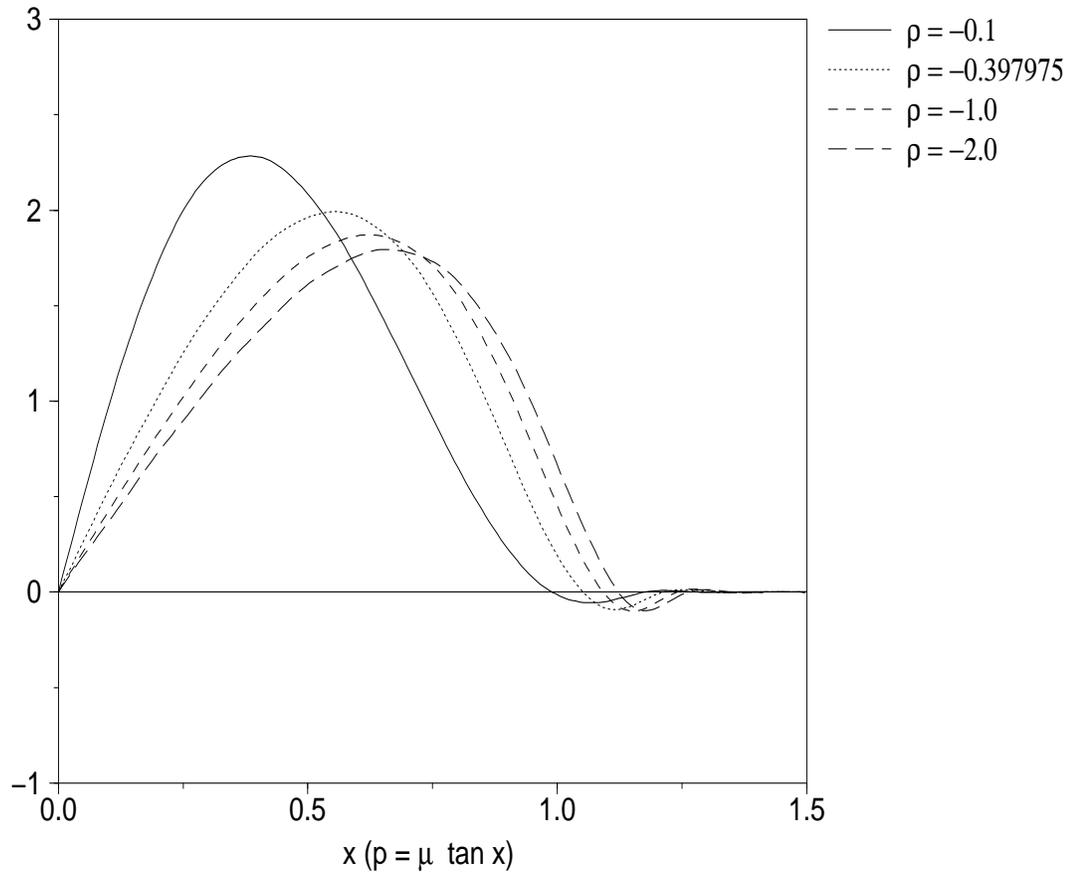,height=12cm,width=12cm,angle=0}
\caption{Wave function (relativistic) of the ground state in momentum
space for the Saxon Woods potential with $m = 1.0$, $R = 10.0$, $a = 1.0$.}
\label{saxwoodpwv}
\end{center}
\end{figure}

 
This collapse is typical for potentials that are not bounded from
below, like the Coulomb and Yukawa potential. To show the difference,
we have done the same calculations for the Saxon-Woods potential
\begin{eqnarray}
V(r) = \frac{\rho}{1 + e^{(r - R)/a}}. 
\end{eqnarray}
This potential is bounded from below, which means that the bound state
energies cannot drop to infinity. We have calculated the bound state
energies and wave functions in both configuration and momentum space
for the case
\begin{equation}
m = 1.0, R = 10.0, a = 1.0, \mu = 0.35
\end{equation}
and several values of $\rho$. The bound state energies are given in
table~\ref{saxwood}, in Fig.~\ref{saxwoodrwv} the wave functions in
configuration space are plotted and in Fig.~\ref{saxwoodpwv} they are
plotted in the momentum space representation.
Looking at Figs.~\ref{saxwoodrwv} and \ref{saxwoodpwv} we see that the
wave functions do not show any irregularities.

\begin{table}[h]
\begin{center}
\caption{Bound state energies for the Saxon Woods potential in the
relativistic case for $m = 1.0$, $R = 10.0$, $a = 1.0$. \label{saxwood}}

\vspace{1ex}

\begin{tabular}{|c|c|c|c|c|}
\hline
n & $\rho = -0.1$ & $\rho = -0.397975$ & $\rho = -1.0$ & $\rho = -2.0$ \\
\hline
0 & -0.04328906536 & -0.3107855320 & -0.8903754950 & -1.870764375 \\
1 & -- & -0.1143078804 & -0.6332213879 & -1.569842815 \\
2 & -- & -- & -0.3176877499 & -1.184906960 \\
3 & -- & -- & -0.03023195267 & -0.7476220131 \\
\hline
\end{tabular}
\end{center}
\end{table}


 
Finally we have plotted the ground state energy versus the strength
$\rho$ for the Saxon Woods potential in the same figure with the Yukawa
potential (Fig.~\ref{rhoenergy}). Comparing these two potentials, we
notice that the curve of the Saxon Woods potential goes down smoothly
instead of dropping down, as for the Yukawa
potential. The reason for this behaviour is that the Saxon Woods
potential is bounded from below and does not have a collapse.   

\section{Conclusion}

The easiest way to obtain the bound states is to directly solve
the differential equation. This method is very easy to implement
and calculations do not take up much time. Unfortunately it is
only usefull in the nonrelativistic situation. Beside, it is not
very accurate.

In the relativistic situation we use two other methods; discretizing
the integral equation and making an expansion in terms of basis
functions. For both methods the implementation is more complicated
than in the case of directly solving the differential equation,
but it is not extremely difficult. Calculations do take up some more
time, but this is negligible. And both methods are more accurate than
the method of directly solving the differential equation.

A difference between the methods is that the calculations with the
method of discretizing the integral equation are done in momentum
space only. When making an expansion in terms of basis functions the
kinetic and potential energies are seperately calculated in either
momentum or configuration space. A consequence of this difference is
that the method of discretizing the integral equation is not very
wellsuited to all potentials, since we may have to deal with
singularities in the potential matrix elements.\\

After checking our codes with the exact solutions of several potentials,
we have calculated the bound state energies and wave functions for
the Yukawa potential in the relativistic case. We found a more deeply
bound state in the relativistic situation compared to the nonrelativistic
one. We have also looked at what happens to the wave functions when
the strength of the potential increases. It turned out that there
exists a critical value for the strength above which the wave functions
collapse. This collapse only occurs for potentials that are not
bounded from below; e.g. the Coulomb or Yukawa potential. Potentials
that are bounded from below do not show this collapse, since their
ground state energies do not drop down to minus infinity. As an example
of this kind of potential we have looked at the Saxon Woods potential.

\end{document}